# Measuring Asymmetric Opinions on Online Social Interrelationship with Language and Network Features

Bo Wang
Tianjin University

Yanshu Yu
Tianjin University

Yuan Wang
Tianjin University

## ABSTRACT
Instead of studying the properties of social relationship from an objective view, in this paper, we focus on individuals' subjective and asymmetric opinions on their interrelationships. Inspired by the theories from sociolinguistics, we investigate two individuals' opinions on their interrelationship with their interactive language features. Eliminating the difference of personal language style, we clarify that the asymmetry of interactive language feature values can indicate individuals' asymmetric opinions on their interrelationship. We also discuss how the degree of opinions' asymmetry is related to the individual's personality traits. Furthermore, to measure the individuals' asymmetric opinions on interrelationship concretely, we develop a novel model synthetizing interactive language and social network features. The experimental results with Enron email dataset provide multiple evidences of the asymmetric opinions on interrelationship, and also verify the effectiveness of the proposed model in measuring the degree of opinions' asymmetry.

## 1. INTRODUCTION
The main problem that we are attempting to investigate in this paper is the following: given a pair of individuals who are engaged in discourse, are their subjective opinions of their interrelationship symmetric or not, and how can we measure their subjective opinions on their interrelationship?

To recognize the nature of interrelationships between individuals is a very essential problem in social computing, which is the basis of many related studies such as community discovery, influence analysis, link predication and recommendation [3]. Actually, when we try to determine the nature of an interrelationship, we have two primary options: The first one is regarding the nature of interrelationships as objective properties, which can be investigated independently of the participants' subjective opinions. The second option is to determine the nature of interrelationships according to two participants' subjective opinions on their interrelationships. In practice, the objective measurement often leads to symmetric values of the interrelationships' properties, while the subjective measurement can leads to asymmetric values, because two participants may have different opinions on their interrelationship. The verification and measurement of this kind of subjective asymmetry is important in many social computing studies, For example, in influence analysis, the information propagation possibility is not balanced between two individuals: the one who believes their interrelationship is strong tend to pass/receive more information, and the other one who believes the interrelationship is weak will do the opposite.

Generally speaking, the important problems of the subjective measurement of interrelationship should include: (1) Can we verify the intuition about the opinions' asymmetry on social interrelationship with experimental evidence? (2) How to measure the asymmetric opinions on interrelationship? (3) Does the degree of the subjective asymmetry of the interrelationship has any latent reasons, e.g., the personality traits of the participant individuals?

To answer these questions, the key is to find a more direct method to investigate the individuals' subjective opinions on their interrelationship. In this work, we propose to do this with the features from interactive language between individuals. In section 3.2, according to related theories in sociolinguistics, we select four typical interactive language features which indicate the frequency, quantity, quality and emotion of interactive language, respectively. In section 3.3 we further normalize these linguistic features with each individual's preference of language using to distinguish the personal opinions from the personal language habit. In section 4, we propose an optimization process to measure the individuals' asymmetric opinions of their interrelationship with their interactive language features. The experiments in section 5 investigate the asymmetry of individuals' opinions on interrelationship with multiple experimental evidences. The experiments in section 6 verify the effectiveness of the proposed measurement of the asymmetric opinions on interrelationship.

## 2. RELATED WORK
### 2.1 Symmetric & Asymmetric Analysis of Signed Social Interrelationships
Many early studies of the social relationship suppose the properties of interrelationships are symmetric. Among these works, some studies work on unsigned relationship and recently researchers pay more attention to the signed relationship analysis in which the interrelationships are labeled to illustrate typical types such as positive and negative.

The problem of predicting edge signs in a social network was first considered by Guha et al. [13]. They developed a trust propagation framework to predict the trust (or distrust) between pairs of nodes. Many recent studies focused on the positive/negative analysis [4, 6, 11]. Kunegis et al. [18] proved the added value of negative links in social networks. They also studied the spectral properties of signed networks and used kernels derived from a signed variant of the graph Laplacian for link prediction [19]. Brzozowski et al. [10] applied random walk algorithms to propagate both positive and negative links on Epinion network. Kunegis et al. [17] conducted similar experiments on Slashdot data.

There are also some latest studies analyzed the social relationship directionally or asymmetrically in a sense. For directed relationships, Leskovec et al. [20] first considered an explicit formulation of the sign prediction problem. Their prediction methods are based on the theory of social balance and status. Bach et al. [5] and Huang et al. [15] framed sign prediction as a hinge-loss Markov random field, a type of probabilistic graphical model introduced by Broecheler et al. [9]. West et al. [27] developed a model that synthesizes textual and social-network information to jointly predict the polarity of person-to-person evaluations. This work focused on the mutual evaluation between two individuals which are naturally asymmetric and was an important enlightenment of our work. Different from this work, we focus on the bidirectional evaluation of interrelationship.

## 2.2 Synthesis of Language Analysis & Social Network Analysis

Natural language processing and signed social networks analysis have been successful in a variety of tasks. In recent studies, these two technologies are often combined to understand the nature of the social relationship, user profile or social event. Adalı et al. [1] showed that the feature sets from social behavioral information and textual information are practically equivalent in terms of their ability to determine the different types of relationships between pairs of individuals interacting in social media. Bramsen et al. [8] presented a text corpus-based statistical learning approach to modeling social power relationships. Thomas et al. [25] and Tan et al. [26] proposed a graph-cuts approach of Pang [23]. Ma et al. [22] and Hu et al. [15] enforced homophily between friends with regard to their preferences.

However, compared with state-of-the-arts works, we focus on the bidirectional evaluation of the interrelationship. For two individuals, their mutual evaluation of each other is not always equivalent to their evaluation of their interrelationship. Though previous works did not analyze the subjective opinions on interrelationships asymmetrically, they are good references of our work encouraging us to apply various analyses of the language content in the social network to analyze the interrelationship with subjective information from the languages. In this work, we explore the possibility to understand the asymmetric opinion on interrelationship with interactive language features and develop a model to measure individuals' asymmetric opinions on their interrelationships.

## 3. CHARACTERIZE OPINIONS ON INTERRELATIONSHIP WITH INTERACTIVE LANGUAGE FEATURES

In current studies, when researchers want to measure the strength or type (e.g., positive & negative) of social interrelationship between two individuals from an objective view, they often turn to typical objective features, e.g., the embeddedness of the interrelationship or the communication frequency between the participants. The results of such methods are often symmetrical i.e., independent of the direction of interrelationship. While, if we suppose the properties of interrelationship are not symmetrical, i.e., are different for two directions, it is often induced by the difference of the opinions from the two participants. Then, the problem should be how can we describe the individuals' opinions on their interrelationship asymmetrically? Some studies choose to use objective features directionally, e.g., the communication frequency in two directions. Though some objective features can be used asymmetrically, they cannot describe the opinions accurately. Therefore, a more reasonable solution is to measure the opinions on interrelationship with subjective features.

Among the available resources, interactive language is a good choice, because the interactive language is not only closely related to the properties of interrelationship, but also highly descriptive of the opinions of individuals. In this section, we compare the objective and subjective features of interrelationship and then investigate the opinions of interrelationship asymmetrically with several typical subjective features of interactive language, extending the investigation in [7].

### 3.1 Objective Features for Interrelationship

Most popular objective features to measure the social relationships are the embeddedness of an interrelationship and the communication frequency between two participants.

The embeddedness can measure the strength of an interrelationship with the coincidence degree of two participants' interrelationship, i.e., the count of the common "friends" in many cases. The embeddedness is an effective objective feature which describes the interrelationship with only the structure of the social network, but it is not designed to be measured directionally.

Communication frequency can measure the strength of an interrelationship with the frequency of certain type of communication between two participants. The most widely used cases include the frequency of phone call, email and mutual comments on social media. The communication frequency is not an absolute objective feature, because it is actually a kind of behavior of the participants and is a subjective decision. Though the subjective property makes the communication frequency can naturally be measured directionally, it can only reflect the opinions indirectly because of the limited information. For example, in some cases, though one have similar email frequency with two colleagues, he may have totally different attitudes to his interrelationships with these two colleagues.

Compared with each other, during the interrelationship analysis, the advantage of objective features is the convenience to be quantified and measured, while the advantage of subjective features is the ability to recognize the participants' opinions on interrelationship more directly and accurately. For the above example, though one have similar email frequency with two colleagues, his different attitudes towards two interrelationships can still be revealed by the content of his emails. Therefore, to measure the opinions, subjective features from interactive language can be more helpful than the objective features.

### 3.2 Sociolinguistics and Subjective Features

Subjective features are the features which can reveal the participants' subjective opinions on their interrelationships. In sociolinguistics, the theory of communicative action [14] proposes to reconstruct the concept of relationship with the communicative act, instead of the objectivistic terms. And the linguistic structures of communication can be used to establish a normative understanding of the social relationships. Sapir-Whorf hypothesis [24] also supposes that the semantic structure of the language using shapes or limits the ways in which a speaker forms conceptions of the world including the social relationships. In the communication, people's choice of words is always highly depending on their subjective opinions on their interrelationship with the others. These theories inspire us to make the assumption that one's opinion on an interrelationship can impact his choice of language style in communication. Consequently, we can investigate one's opinion on his interrelationships according to his interactive language style on his interrelationships.

The next problem is how to measure an individual's interactive language style. In sociolinguistics, Holmes [16] introduced four important dimensions to study the language using in the social communication:

(1) The solidarity-social distance scale: concerned the solidarity of the individuals' relationship in social communication.

(2) The social status scale: concerned the relative status of the individuals' relationship in social communication.

(3) The formality scale: concerned the formality of language using in different relationships, topics and places.

(4) The referential and affective scale: concerned referential and affective function of the language in social communication.

The first two dimensions concerned the features of social interrelationship from both subjective and objective views. The last two dimensions concerned the features of interactive languages, which are highly related to social interrelationship.

Inspired by Holmes' theory, in this work, we propose four typical features of interactive languages to investigate an individual's subjective opinion on his interrelationship. The features are frequency, length, fluency and sentiment polarity which indicate quantity, quality and emotion of interactive language, respectively. Among the features, the frequency and length are two primary features of language communication, while the fluency and sentiment corresponds to the formality and affective scale mentioned in Holmes' theory. It is noted that all these four features can be recognized by the state-of-the-arts natural language processing technologies:

(1) The frequency is the number of times of communication within a specified period of time. In interpersonal communication, one's frequency consists of the passive replay to the other people and the active communication launched by himself. Therefore, frequency can partially reflect one's intention of the communication on an interrelationship.

(2) The length is the number of the words of the interactive language. In interpersonal communication, the length can also partially reveal one's intention of the communication.

(3) The fluency can reflect the formality and quality of the interactive language which is related the use of the words and grammar. In interpersonal communication, the fluency can reveal whether the speaker treats the interrelationship seriously or not which can be influenced by his opinion on the interrelationship.

(4) The sentiment polarity measures the emotion tendency of interactive language. In intuition, the sentiment polarity tends to have a positive correlation with the personal opinion on interrelationship, i.e., more positive emotion in interactive language often indicate that the interrelationship is more valuable to the speaker.

## 3.3 Distinguish the Asymmetry of Opinions from the Asymmetry of Language Habits

Though one's interactive language style is closely related to his opinions on his interrelationships, it is inexact to understand the opinions using the original interactive language features directly. In natural language understanding, the meaning of the language is always not only determined by the content, but also by the context.

The context of the interactive language is very complex including the language habit of the speaker, the occasion of the dialogue, the sentences in the same discourse, and so on. In this work, we focus on how to understand one's opinion more accurately considering his personal habit in language using. For example, suppose $A$ is a very negative people, and he always talks to person $B$ with negative sentiment polarity score. But, we also know that $A$ talks to every people very negatively and he talk to $B$ most friendly compared with the others. In this case, if we want to measure the $A$'s opinion on his relationship with B correctly using sentiment polarity score, we need to calculate the score relatively to $A$'s personal language habit, instead of using the original score.

To distinguish one's opinion from his personal language habit, in this work, for an individual $I$, to characterize $I$'s opinion avoiding the interference from $I$'s personal language habit, we normalize $I$'s language feature score $f$ with $I$'s personal language habit value $H_f(I)$. $H_f(I)$ is measured by Formula (1), where $f(I, I_i)$ is the $f$'s value of the languages said by $I$ to another individual $I_i$, and $C$ is the set of all individuals who is in communication with $I$.

$$H_f(I) = \frac{1}{|C|}\sum_{I_i \in C} f(I, I_i) \quad (1)$$

Then $f'(I, I_i)$ is the normalized value of $f(I, I_i)$ according to $I$'s personal language habit, which can be calculated with Formula (2):

$$f'(I, I_i) = \frac{f(I, I_i) - H_f(I)}{H_f(I)} \quad (2)$$

## 4. MEASURE THE ASYMMETRIC OPINIONS ON INTERRELATIONSHIP WITH SYNTHETIZED FEATURES

In this section, we formulate a computational model to measure the bidirectional degree of individuals' asymmetric opinions on their interrelationship. The model synthesize the proposed interactive language features and the triangle balance of the social interrelationships in social network.

### 4.1 Theory of Social Balance and Extension

Theory of social balance [2] is very popular in social network studies. This theory is based on simple intuitions like 'a friend of my friend is my friend', 'an enemy of my enemy is my friend', and 'an enemy of my friend is my enemy'. In principle, the balance theory is based on the homogeneity assumption. The assumption states that, in a social triangle, the more similar two individuals' opinions on the third one are, the more positive their interrelationship will be, and vice versa. If we sign interrelationships with undirected binary sign '+' and '-', these are statements about the edge signs of triangles according to balance theory: given the signs of two edges, balance theory predicts the third one, as summarized in Figure 1(a), where the two given edges (gray) determine the third one (black).

In this work, to meet the demand of asymmetric opinions measuring, we extend the traditional balance theory to directed triangles. The extended version keep the principle of homogeneity. Figure 1(b) illustrates the extended theory. In Figure 1(b), if individual $A$ and $B$ have similar opinions on their interrelationship with individual $C$, they will all have more positive opinion on their interrelationship from their own point of view.

It is noticeable that, in the extended balance theory, though the signs of the edges are binary and symmetric, these signs are determined by minimizing the cost according to the real-valued language feature in the optimization model. Therefore, the bidirectional symmetry in extended balance theory is not contract to asymmetric modeling of opinions. Actually, the final measurement is a joint result considering both the asymmetric subjective features (language features) and the symmetric objective features (extended balance theory).

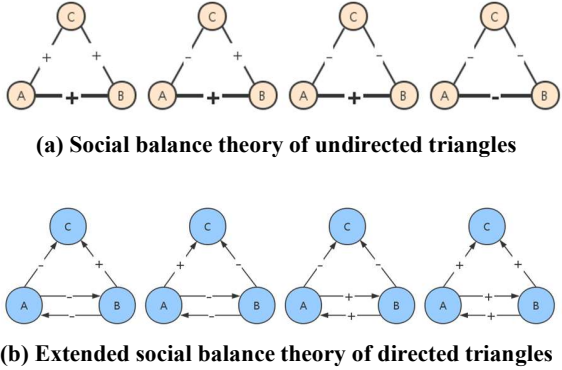

(a) Social balance theory of undirected triangles

(b) Extended social balance theory of directed triangles

**Figure 1. Traditional and extended social balance theory on undirected and directed triangles.**

## 4.2 Optimization Problem

Following the method in [27], in this work, a social network is defined as $G = (V, E, s)$, where $V$ represents individuals; $E$ represents directed relationships; and the vectors set $s \in [0,1]^{|E|}$ represents the subjective strength values on each interrelationship in $E$. We define a directed triangle $t = \{e_{AB}, e_{BA}, e_{AC}, e_{BC}\}$ to be a set of four edges in $E$. We use $s_t = (s_{AB}, s_{BA}, s_{AC}, s_{BC})$ to refer to the subjective strength values on triangle $t$. For each directed edge $e$, $f_e$ measures the values of four proposed language features of $e$, i.e., frequency, length, quality and sentiment. The proposed method is an extension (directed) version of the model in [27]. In the new model, we aim to infer bidirectional subjective strength satisfying two targets as far as possible: (1) accord with the predictions of the four language features models, and (2) directed triangles are agree with extended social balance theory. We define an optimization problem to find strength values set $s^*$ using Formula (3)

$$s^* = \underset{s \in [0,1]^{|E|}}{argmin} \sum_{e \in E, f_e} c(s_e, f_e) + \sum_{t \in T} d(s_t) \quad (3)$$

The first term illustrates how much the predicated strength $s_e$ is different from the prediction $f_e$ of the language feature models. The second term illustrates how unbalanced a triangle is. In Formula (3), we use the following cost function:

$$c(s_e, f_e) = \lambda_1 (1-f_e)s_e + \lambda_0 f_e(1-s_e) \quad (4)$$

Here, $\lambda_1, \lambda_o \in R_+$ are parameters allowing for asymmetric costs for higher or lower strength, respectively. The triangle cost for triangle $t$ is $d(s_t)$, which is calculated as the distance between $s_t$ and nearest balanced triangle. $d(s_t)$ makes the triangles similar to the balanced triangle having low cost.

The problem in Formula (3) is NP-hard. We redefine our problem as a hinge-loss Markov random field (HL-MRF) as the method in [27], which is illustrated in Formula (5)

$$\tilde{c}(s_e, f_e) = \lambda_1 ||s_e - f_e||_+ + \lambda_0 ||f_e - s_e||_+ \quad (5)$$

where $||y||_+ = max\{0, y\}$ is the hinge loss. To rewrite $d(s_t)$, for any $s_t \in [0,1]^4$, we can redefine $d(s_t)$ in Formula (6):

$$\tilde{d}(s_t) = \sum_{z \in \{0,1\}^4} F(s_t, z) \quad (6)$$

where,

$$F(s_t, z) = \begin{cases} t_1 * [1 - f(s_t, z)], z \text{ is a balanced triangle} \\ t_2 * f(s_t, z), Other \end{cases} \quad (7)$$

Here, $t_1, t_2 \in R_+$ are tunable parameters, where

$$f(s_t, z) = ||1 - ||s_t - z||_1||_+ \quad (8)$$

where, $||s_t - z||_1 = \sum_{i=1}^{4} |s_i - z_i|$ is the difference between $s_t$ and a directed triangle $z$. Thus, the relaxation of Formula (3) is Formula (9):

$$s^* = \underset{s \in [0,1]^{|E|}}{argmin} \sum_{e \in E, f_e \in \{f1_e, f2_e, f3_e, f4_e\}} \tilde{c}(s_e, f_e) + \sum_{t \in T} \tilde{d}(s_t) \quad (9)$$

## 5. EXPERIMENTAL INVESTIGATION OF OPINIONS' ASYMMETRY ON INTERRELATIONSHIP

In the first set of experiments, we try to investigate whether the individuals' opinions on their interrelationships are asymmetry or not. To make this investigation, on each interrelationship, we compare the individuals' bidirectional language style with language features, and distinguish their opinions from their personal language habit.

## 5.1 Experimental Setup

To investigate the subjective opinions on interrelationship with interactive language features, we utilized the Enron email dataset. This dataset is the collection of emails from 151 Enron employees. Our results are based on the CMU version (May 7, 2015 Version)[1] containing approximately 0.5M unique email messages.

From the original dataset we identified the users in the collection whose email addresses are followed by @enron.com. There were about 0.2M sent/receive user pairs in email messages, and each pair indicates an interrelationship. To make more reliable investigation, we retained only those interrelationships where a minimum of 15 emails were sent in each direction. The filtered set contains 1078 interrelationships between 647 individuals.

For each directed pair of individuals $<I_i, I_j>$, we calculated four linguistic features with the emails sent from $I_i$, to $I_j$:

(1) To calculate the feature "Frequency", we divided the count of the emails sent from $I_i$, to $I_j$ by the days count between the earliest and latest emails sent from $I_i$, to $I_j$.

(2) To calculate the feature "Length", we divided the total words count of the emails sent from $I_i$, to $I_j$ by the count of the emails sent from $I_i$, to $I_j$.

(3) To calculate the feature "Quality", we use the average score of the perplexity of all sentences in the emails sent from $I_i$, to $I_j$. Higher perplexity score means lower language quality. The perplexity score of each sentence is calculated by the SRI language modeling toolkit[2] (SRILM) with Formula (10).

$$perplexity\_score = 10^{(-logprob/(words-oovs+1))} \quad (10)$$

where *prob* is the generating probability of the sentence, *words* and *oovs* are the count of the words and out of vocabulary words in the sentence, respectively.

---

[1] http://www.cs.cmu.edu/~enron/

[2] http://www.speech.sri.com/projects/srilm/

(4) To calculate the feature "Sentiment", we utilized a sentiment dictionary[3] to count the sentiment words in the sentences. The dictionary contains 8945 sentiment words in English Each positive and negative word values 1 and -1, respectively. Then, the sentiment feature value is calculated by dividing the sum of all sentiment words' scores in the emails sent from $I_i$ to $I_j$ by the count of sentences in all emails sent from $I_i$ to $I_j$.

### 5.2 Asymmetry of Language Styles on Social Interrelationships.

In first experiment, to investigate the asymmetry of language styles on interrelationships, we measured the difference of the language feature values (not normalized by Formula (2)) between two individuals in email exchange pairs. For the selected 1078 pairs of individuals, we calculated the paired T-statistic score between the two sequences of their bidirectional language feature values of exchange emails. The results are shown in Table 1. The first column is the average of the feature values of all individuals; the second column is the average of the pair-wise difference of feature values of all individual pairs; the third and fourth column is the *t* value and 95% confidence interval of the paired T-statistic of the difference. The Pearson correlation between the bidirectional feature values sequences is also recorded in the fifth column.

**Table 1. The statistics of the bidirectional language style with language features.**

| Feature | Avg. of all | Avg. of diff. | t | 95% inter. | P cor. |
|---------|-------------|---------------|----|------------|--------|
| Frequency | .2041 | .1324 | 13 | .113-.151 | .69 |
| Length | 770 | 588 | 12 | 492-684 | .40 |
| Quality | 2367 | 1257 | 12 | 1067-1448 | .36 |
| Sentiment | .0099 | .0045 | 23 | .0041-.0048 | .39 |

First, in the results, measured by paired T-statistic, the differences of all four features' values between the paired individuals are significant (> 95%). Furthermore, the absolute degree of the difference is also remarkable compared with the average values of the features, i.e., for four features, the ration between the average difference and average value is 48.59%.

Second, from correlation score, we can see that only bidirectional frequency have relative high correlation compared with the other three features. One possible explanation is that the send/replay relationships between the emails can lead to similar bidirectional email sending frequency between two individuals. But this send/replay relationship has no significant impact on the language style of the emails' content. Therefore, the bidirectional correlations of the other three features are relative low. This could be an evidence which indicates the difference between the objective features (e.g., frequency) and subjective features (e.g., the length, quality and sentiment), and also indicates the necessity to use subjective language features to measure the individuals' real opinions on the interrelationships.

The results in Table 1 provide initial evidence that individuals in email exchange pairs generally have different opinions on their interrelationships, which is indicated by their different language styles in emails' content. But this judgment is not convincing enough, because the interactive language style is not only determined by the opinions on the interrelationships. Another important factor is individuals' personal language habit. In the following experiments, we'll try to distinguish the opinions on interrelationships from the personal language habit.

### 5.3 The Variety of Personal Language Habit

If we want to measure the difference between individuals' opinions on their interrelationships through the interactive language, we have to consider the interference from the personal language habit. Otherwise, we cannot judge whether the difference between the interactive language styles is caused by their different opinions or their different language habit.

Firstly, in Table 2, we investigated whether the personal language habit of individuals are really different or not. For each individual, we calculated the standard deviation of his interactive language feature values toward different communicators, which indicates the range of his language style, i.e., his language habit. Then, we compared the average of all personal deviations and global deviation of all individuals' language feature values. The results show that the average of personal deviations (column 2) is significantly smaller than the global deviation (column 1), which indicates that the ranges of individuals' personal language feature values (i.e., personal language habit) are quite different from each other. This result is also supported by the large deviation among individual's deviations of language feature values (column 3).

**Table 2. The deviations of individuals' ranges of language feature values (personal language habit).**

| Feature | Global Dev. | Individual's Dev. | Dev. of Individual's Dev. |
|---------|-------------|-------------------|---------------------------|
| Frequency | 0.2070 | 0.1093 | 0.0989 |
| Length | 1563 | 346 | 979 |
| Quality | 1368 | 633 | 484 |
| Sentiment | 0.0045 | 0.0024 | 0.0015 |

It's noted that in Table 2, we selected the individuals who have at least five communicators in the data set (112 individuals and 334 interrelationships after selection). This result reveal that the difference of personal language habit is remarkable and it's necessary to distinguish the opinions on interrelationships from the personal language habit.

### 5.4 Asymmetry of Individuals' Opinions on Interrelationship

In this experiment, we measured the asymmetry of the interactive language feature values on interrelationships considering two different causes: personal opinions and personal language habit.

Firstly, we investigated the asymmetry of individuals' language habit on interrelationships. In the Table 3, we calculated the significance of personal language habits difference between individuals. For an individual $I_i$, his language habit value on a language feature *f* denoted by $H_f(I_i)$ is measured by Formula (1).

In Table 3, the first column is the average of all individuals' language habit values on each feature; the second column is the average of the difference between two individuals' language habit values on all interrelationships; the third and fourth column is the *t* value and 95% confidence interval of the paired T-statistic of the bidirectional personal habit values of all pairs of individuals' in communication. The Pearson correlation between the bidirectional personal habit values is also recorded in the fifth column.

---
[3] http://www.keenage.com/download/sentiment.rar

In Table 3, the asymmetry (difference) of the personal language habits on interrelationships are significant (>95%) on all four features. Then, can this result indicate that the asymmetry of the language features shown in Table 1 is only caused by the difference between personal language habits, and is not an evidence of the asymmetry of opinions on interrelationships? The next experiment will answer this question.

**Table 3. The statistics of the bidirectional individuals' language habits on interrelationships.**

| Feature | Avg. of all | Avg. of diff. | t | 95% inter. | P cor. |
|---|---|---|---|---|---|
| Frequency | 0.204 | 0.1289 | 17 | 0.114-.143 | 0.70 |
| Length | 770 | 542 | 16 | 477-608 | 0.57 |
| Quality | 2367 | 1185 | 12 | 1003-1368 | 0.32 |
| Sentiment | 0.01 | 0.0041 | 23 | 0.0037-.0044 | 0.35 |

In the next experiment, we investigated the asymmetry of individuals' opinions on interrelationships. We normalized the features of language style by individuals' language habit with Formula (2). The normalized values can avoid the interference of personal language habit, and thus reveal individuals' actual opinions on their interrelationships. The results are shown in Table 4, where the first column is the average of the individuals' opinions values of the interrelationship (i.e., the normalized language features values); the second column is the average of the difference between two individuals' opinion values on all interrelationships; the third and fourth columns are the *t* value and 95% confidence interval of the paired T-statistic of the bidirectional opinions values on interrelationships. The Pearson correlation between the bidirectional opinions values is also recorded in the fifth column.

**Table 4. The statistics of the bidirectional individuals' opinions on interrelationships.**

| Feature | Avg. of all | Avg. of diff. | t | 95% inter. | P cor. |
|---|---|---|---|---|---|
| Frequency | 0.383 | 0.3380 | 22 | 0.3082-0.3677 | 0.53 |
| Length | 0.443 | 0.3572 | 25 | 0.3296-0.3848 | 0.50 |
| Quality | 0.434 | 0.3614 | 24 | 0.3320-0.3909 | 0.51 |
| Sentiment | 0.425 | 0.3374 | 23 | 0.3094-0.3655 | 0.48 |

As shown in Table 4, avoiding the interference from the personal language habit, the difference of the language feature values on interrelationships are still significant (>95%) on all four features, which is a more reliable evidence of the asymmetry of personal opinions on interrelationships.

With the experimental results in Table 3 and 4, we can see that the asymmetry of language feature values on interrelationship is a joint result caused by the asymmetry of personal language habit and the asymmetry of personal opinions on interrelationship. It is also noted that, the greater *t* values in Table 4 than that in Table 3 indicates that on interrelationships, the asymmetry of personal opinions is more significant than language habit.

### 5.5 The Role of Personality Traits

In this experiment, we further tried to investigate possible latent reasons of the opinions' asymmetry on social interrelationship. We focus on the personality traits of the individuals.

Here we consider two kinds of personality traits: the positive degree and the flexible degree. An individual has higher positive degree means that he adopts more positive attitude in social relationship, and higher flexible degree means that his social behavior is more changeable instead of adhering to a fixed style.

We supposed that the positive degree of an individual can be measured by his language feature values, i.e., higher positive degree can be indicated by higher scores of frequency, length, quality, and sentiment. We also supposed that the flexible degree can be measured by the standard deviation of his language feature values, i.e., higher flexible degree can be indicated by larger deviation of the language feature values.

With above assumption, for each language feature, we selected the individuals having the top/last 10 average/deviation values on language features, who are the individuals having the highest/lowest positive/flexible degree, respectively.

For an individual *I*, *I*'s 'correlation score' in this experiment denotes how well *I*'s language feature values correlate the language feature values of *I*'s partners in *I*'s interrelationships. A better 'correlation score' means *I*'s language style match the partners' language styles better on interrelationships, and consequently the degree of the opinions' asymmetry on *I*'s interrelationships is lower.

In Figure 2, for top/last 10 positive/flexible individuals on each language feature, we calculated the average of ten individuals' correlation scores. The X-axis is the language features and the Y-axis is the average correlation score. It can be seen that the individuals having the top10 positive/flexible personality obtain better correlation scores than the individuals having the last10 positive/flexible personality on all four features.

Given the above assumption, this result indicates that, on this dataset, the individuals who have more positive and flexible personality traits tend to adapt to other people's style in the communication and thus lead to less asymmetry of the opinions on interrelationships. This experimental result inspires us to understand and predicate the asymmetry of the opinions on interrelationship with the help of individuals' personality traits.

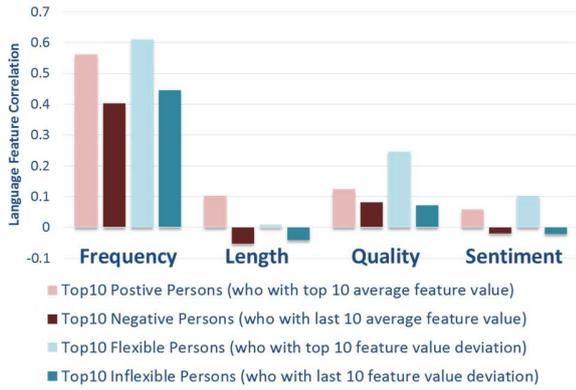

**Figure. 2. Bidirectional language feature correlations on the interrelationships of persons with different personality.**

## 6. EXPERIMENTS OF MEASURING ASYMMETRIC OPINIONS ON INTERRELATIONSHIP

In this experiment, we evaluate the proposed model's ability to measure the bidirectional degrees of individuals' asymmetric opinions on their interrelationships. We describe used data and ground truth based on Enron email dataset, and then present the experimental results to compare the performance of the proposed synthetized model and the performance of single features models.

### 6.1 Data Setting and Ground Truth
**Data Setting:**

For this experiment, in the previous filtered set which contains 1078 interrelationships between 647 individuals, we obtain the organizational hierarchy of 232 Enron employees provided by Apoorv Agarwal [2]. With this dataset, we can determine the superior-subordinate relationships between individuals. Finally, 70 pairs of the superior-subordinate relationships are manually exploited consists of 80 nodes and 140 directed edges. We conducted experiments with the proposed model (Formula (9)) synthetizing normalized pair-wise language features (Formula (2)) and directed triangles of interrelationships.

**Ground truth:**

To find the ground truth about individual's exact opinions on their interrelationship is difficult. On this email dataset, we intuitively make the assumption that in a superior-subordinate relationship, two participants' opinion on their interrelationship tend to be asymmetric. The individual of lower position tend to put more importance on their interrelationship than the individual of higher position. We use this assumption as the ground truth to evaluate the proposed models in measuring asymmetric opinions.

### 6.2 Performance Comparison
In the experiment, if a pair of individuals *A* and *B* are in a superior-subordinate relationship and *A* is the one of lower position. *Score_A* and *score_B* are the degree of *A* and *B*'s opinions on their interrelationship mesaured by the proposed model, respectively. If *score_A – score_B > threshold*, we argue that the asymmetric opinions measurement on this pair is successful, i.e., the individual of lower position (*A*) put more importance on their interrelationship than the individual of higher position (*B*). In experiment, we tried the threshold = 0, 0.01, 0.05, 0.1, respectively. The chosen thresholds are zero and three nonzero values equally spaced from 0.01 to 0.1. The threshold is supposed to tune the result slightly. A too large threshold will bring unnecessary manual influence to the model.

In comparison, the results measured by single language features are noted by the names of features, i.e., 'Frequency', 'Length', 'Quality' and 'Sentiment'; the result of proposed model synthetizing only four language features is noted as 'Language_Features'; the proposed model synthetizing language and social network (triangle balance) features is noted as 'Language_Features+Structural_Feature'.

The precision of the asymmetric opinions measurement are shown in Figure 3. The X-axis is the value of thresholds and the Y-axis is the precision of opinion measurement.

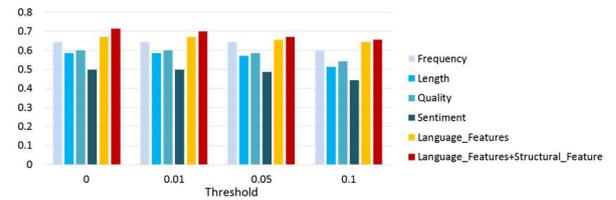

**Figure 3. The performance of single and synthetized models in measuring the asymmetric opinions on interrelationship between superior-subordinate user pairs in Enron email.**

As shown in Figure 3, the 'Language_Features' model performs better than all single features on all thresholds. When synthetized with the structural feature, the performance of 'Language_Features+Structural_Feature' model surpasses the 'Language_Features' model on all thresholds. This result is a case study verifying the effectiveness of the proposed model in measuring the asymmetric opinions on social interrelationship. The key advantage is to synthetize the features from interactive language and social network structures.

## 7. CONCLUSIONS
In this paper, we investigated and measured the asymmetry of individuals' opinions on social interrelationship, with the help of interactive language features. According to sociolinguistics theories, we adopted four key language features to represent one's interactive language style, including the frequency, length, quality and sentiment.

As a case study, with these features, we investigated individuals' opinions on interrelationships with their exchange email contents in Enron email dataset. The experimental results indicated that the asymmetry of individual's language styles is a joint effect of the asymmetry of their opinions and the asymmetry of their language habits. We also found that individuals' personality traits can be a latent cause of their asymmetric opinions on interrelationship: positive and flexible personality may alleviate the opinions' asymmetry on interrelationships.

We further developed a model that synthesizes interactive language and social network features to jointly measure the concrete degree of asymmetric opinions on social interrelationship. Using superior-subordinate relationship among Enron email users as ground truth, as a case study, we initially verified the effectiveness of the proposed model in measuring asymmetric opinions on interrelationship. In the experiments, the synthesized model outperformed the baselines using combined or single language features.

Our experimental research provides initial evidences and measurement of the individuals' asymmetric opinions on the interrelationships, and potentially leads to a promising direction to model the social relationship from a subjective view. In the future work, we will try to describe the opinion on interrelationship with more detailed information using the content of interactive language.